\documentclass[rnote]{aa} 
\usepackage{graphicx}
\usepackage{txfonts}
\begin{document}

\title{Another deep dimming of the classical T Tauri star RW Aur A
          \thanks{Based on observations collected at the Nordic Optical Telescope, La Palma, Spain; Fast-Track Service program 50-409.}
}					
   

   \author{ P. P. Petrov\inst{1}
                       \and G. F. Gahm\inst{2}
                       \and A. A. Djupvik\inst{3}
		\and E. V. Babina\inst{1}
		\and S. A. Artemenko\inst{1}
		\and K. N. Grankin \inst{1}
           }
           
   \offprints{P. P. Petrov}

     \institute{Crimean Astrophysical Observatory, p/o Nauchny, 298409 Republic of Crimea\\
              email: \mbox{petrov@crao.crimea.ua; petrogen@rambler.ru}\\
              \and Stockholm Observatory, AlbaNova University Centre, Stockholm University, SE-106 91 Stockholm, Sweden \\
              \and Nordic Optical Telescope, Rambla Jos\'{e} Ana Fern\'{a}ndez P\'{e}rez 7, ES-38711 Bre\~{n}a Baja, Spain
	  }
	  
   \date{}
   
   
\abstract
	 {RW Aur A is a classical T Tauri star (CTTS) with an unusually rich emission line spectrum. In 2014 the star faded by $\sim$3 magnitudes in the $V$ band and went into a long-lasting minimum. In 2010 the star suffered from a similar fading, although less deep. These events in RW Aur A are very unusual among the CTTS, and have been attributed to occultations by passing dust clouds. }
   {We want to find out if any spectral changes took place after the last fading of RW Aur A with the intention to gather more information on the occulting body and the cause of the phenomenon. }
   {We collected spectra of the two components of RW Aur.  Photometry was made before and during the minimum.}
   { The overall spectral signatures reflecting emission from accretion flows from disk to star did not change after the fading. However, blue-shifted absorption components related to the stellar wind had increased in strength in certain resonance lines, and the profiles and strengths, but not fluxes, of forbidden lines had become drastically different. }
  {The extinction through the obscuring cloud is grey indicating the presence of large dust grains. At the same time, there are no traces of related absorbing gas. The cloud occults the star and the interior part of the stellar wind, but not the wind/jet further out. The dimming in 2014 was not accompanied by changes in the accretion flows at the stellar surface. There is evidence that the structure and velocity pattern of the stellar wind did change significantly. The dimmings could be related to passing condensations in a tidally disrupted disk, as proposed earlier, but we also speculate that large dust grains have been stirred up from the inclined disk into the line-of-sight through the interaction with an enhanced wind. } 
	   
 \keywords{stars: pre-main sequence -- stars: variables: T Tau -- stars: circumstellar matter -- stars: individual: RW Aur}

 \maketitle


\section{Introduction}
\label{sec:intro}

The classical T Tauri star (CTTS) RW Aur A stands out among the low-mass pre-main-sequence (PMS) stars with its unusually rich emission line spectrum as noted early by Joy (\cite{Joy45}). The star has been subject to a large number of investigations. Several spectral features respond to changes in the accretion rate, like narrow components in emission lines of e.g. \ion{He}{i}, excess continuous emission (called veiling), and red-shifted absorption components flanking certain strong emission lines. These signatures are prominent in RW~Aur A and indicate that the accretion from a circumstellar disk is heavy and variable (e.g., Petrov et al. \cite{Petrov01a}; hereafter called P2001). RW Aur is a visual binary with a separation of 1.4$\arcsec$, where the faint component B is a weak-line TTS, a PMS star with little or no evidence of accretion. 

In 2010, RW Aur A went through a deep minimum reaching an amplitude of $\sim$2 magnitudes in $V$ and which lasted for 180 days  (Rodriguez et al. \cite{Rodriguez13}; hereafter called R2013). They attributed this drop to an occultation by part of a tidally disrupted disk as evidenced in Cabrit et al. (\cite{Cabrit06}) and noted that such a long-lasting event had never been observed before in at least 50 years. Chou et al. (\cite{Chou13}) collected high-resolution spectra during the beginning of the minimum indicating that no significant changes in the emission line spectrum had occurred.

Normally, the star fluctuates in brightness by sometimes more than one magnitude in the $V$ band and on time scales of a few days. No distinct period has been found as summarised in e.g. Gahm et al. (1993) and R2013.  However, colour changes with periods of 2.7 to 2.8 days have been reported (Petrov et al. \cite{Petrov01b}, R2013), which is close to half the expected rotational period of about 5.6 days as determined from variations  of the longitudal magnetic field of the star (Dodin et al. \cite{Dodin12}). The star becomes redder with decreasing brightness, and the frequent drops in brightness from an average level of $V \approx 10.4$ appears to be, at least in part,  related to variable foreground extinction (e.g., Herbst et al. \cite{Herbst94}). 
 
In 2014, RW Aur A entered a second long-lasting minimum in brightness, and this time even deeper than in 2010. According to the data collected by the American Association of Variable Star Observers (AAVSO) the brightness dropped by more than 2 magnitudes between May 1 and October 23. Resolved $UBVRI$ photometry of both components of RW Aur was performed on November 13/14 2014 by Antipin et al. (\cite{Antipin15}). They found that RW Aur A was $\sim$3 magnitudes fainter in all bands compared to normal levels and that the star was fainter than component B on this date. Furthermore, they concluded that the drop in brightness was caused by an increase in foreground, mainly grey extinction.

The long-lasting minima of RW Aur A are very rare among the CTTS. A special case is KH 15D, where binary components in an inclined orbit repeatedly dives behind the disk (Hamilton et al. \cite{Hamilton12}). Another case is AA Tau, a CTTS with a warped disk seen almost edge-on. This star suffered from a deep extinction in 2011 and has remained in this state, while the accretion rate has not changed by much (Bouvier et al. {\cite{Bouvier13}). Unlike AA Tau, the disk of RW Aur A is inclined by between 45 to 60 degrees as estimated in Cabrit et al. (\cite{Cabrit06}). We note that a few other weak-line TTS have shown more extended eclipses by foreground dust (see e.g., Grinin et al. \cite{Grinin08}; ; Mamajek et al. \cite{Mamajek12}). 

On the other hand, a subclass of early-type PMS stars called UX Ori stars (see Waters \& Waelkens \cite{Waters98}) are characterized by their long lasting eclipses. These events are recurrent and cyclic in most cases, which was one reason that led R2013 to conclude that the unique dimming of RW Aur A was of a different origin than in the UX Ori stars, where inhomogeneities in remote parts of surrounding disks or envelopes are thought be the cause (e.g. Grinin et al. \cite{Grinin98}; Grinin \& Tambovtseva \cite{Grinin02}).  

Since the disk in RW Aur A is tilted to the line-of-sight it follows that if the occulting material is distant from the star it must reside at some height above the disk plane. In this case spectral signatures of a disk wind, as manifested in certain forbidden lines, may have changed after the fading. On the other hand, if the occulting body is closer to the star, at the inner disk edge, the fading might be accompanied by changes in the accretion signatures in the spectrum of RW Aur A. Also, if the fading is associated to changes in the strength or configuration  of the stellar wind this may be traced in the blue-shifted part of certain emission lines and as changes in forbidden lines forming in the wind. Finally, absorption lines from the gas in the occulting body could appear in the spectrum.

In order to search for any spectral changes that could be associated with the fading of RW Aur A we collected spectra in November and December 2014 covering the optical spectral region. Photometric backup was also provided. In the present Reseach Note we present our findings from this material.

\section{Observations}
\label{sec:observation}

Observing time was granted via the Nordic Optical Telescope (NOT), and two spectra of RW Aur were taken on December 23 (UT 00:38) and Dec. 26 (UT 21:58), 2014 using ALFOSC. The first spectrum was obtained with the high efficiency VPH grism \#17 with a 0.5'' wide slit to obtain a spectral resolution ($\lambda/\Delta \lambda$) of 10000 around H$_{\alpha}$ (6350-6850 $\AA $). The second spectrum was obtained with the echelle setup of grism \#9, using grism \#10 as a cross-disperser, and a 1.0'' wide echelle slitlet, to obtain a coverage from 3300-10350 $\AA $ with a spectral resolution around 2000. In both cases, the slits were oriented to include both components of the 1.4'' separation binary, and the seeing conditions were adequate (FWHM $\leq$ 1.0'') to secure spectra of both components A and B. The exposure times were chosen to give a S/N ratio of 100 in the continuum for the latest magnitude measurement of component A. For the second setup we also observed the standard star HD19445. Halogen flats and arc lamp exposures were obtained at the pointing of the telescope to facilitate fringe removal and minimize the impact of instrument flexure on the wavelength calibration. The spectra were reduced with the IRAF routines, including standard procedures of bias subtraction, spikes removal, flat field correction, and the wavelength calibration using He-Ne and Th-Ar lamps. 

In addition, high-resolution ($\lambda/\Delta \lambda$ =26000) spectra of RW Aur were obtained at the Crimean Astrophysical Observatory (CrAO) on November 3 (UT 02:28 ) and Nov. 14 (UT 00:23) 2014,  using the coud{\'e} echelle spectrograph at  the 2.6 m Shajn reflector. The seeing was not good enough to resolve the binary on the entrance slit, so the registered  spectra include both components. Nevertheless, the spectra were useful to inspect the velocity profile of  forbidden line components that form in the wind of the primary component. Reduction of the CrAO spectra was also done using standard IRAF routines.

The spectroscopic observations were accompanied with photometric patrol at CrAO at the 1.25m telescope (AZT-11) equipped with the Finnish five-channel photometer and the ProLine PL23042 CCD detector. An entrance diaphragm of 15'' was used, i.e. the combined flux from both components was measured. Photometric observations of RW Aur have been carried out in CrAO since 2007.

\section{Results}
\label{sec:results}

The light curve of RW Aur during the period 2010--2014 is shown in Fig.~\ref{curve} with data from our photometric patrol observations. The $V$ versus $(V - R)$ diagram also includes data from Grankin et al. (\cite{Grankin07}) covering a longer period. During the minima in 2010 and 2014 the star dropped far below its normal brightness levels. The two components of RW Aur were not resolved in our photometry and light from the B component contributes when the star is faint. However, in November 2014 Antipin et al. (\cite{Antipin15}) obtained resolved photometry and derived an extinction toward RW Aur A of $A_{V}\sim$3$^m$. They concluded that the extinction could be matched with a totally grey component $A_{V}$ = 2.87$^m$ plus a selective one of $A_{V}$ = 0.44$^m$, which corresponds to the interstellar extinction to RW Aur (e.g. P2001).

In order to find differences between spectra of the dimmed star and those taken at normal brightness we use our collection of high-resolution spectra of RW Aur A obtained in 1996 -1999 at NOT with the Sofin spectrograph (see P2001). The emission line profiles in RW Aur A vary considerably from night to night, and our collection of Sofin spectra is a repesentative sample of this variability. We selected a few Sofin spectra covering the range of the most extreme variations for comparison with our 2014 spectra.

We first note that no changes had occurred in the spectrum of RW Aur B. The spectra of RW Aur A from the normal and faint state are very similar with broad emission lines of Hydrogen, Helium and metals. In particular, the characteristic double-peaked profile of H$\alpha$ remained the same as well as the intensity of the \ion{He}{i} line at 5875 $\AA$, which indicates that no dramatic change has occurred in the accretion rate of gas to the stellar surface during the minimum. This is also reflected by the fact that no change occurred in the degree of veiling, which could be estimated to within 1.5 -- 2.0 (veiling factor) from a few resolved absorption lines of \ion{Ca}{I} and \ion{Li}{I}. The strengths of broad emission lines of metals are somewhat lower than normal, but remain within the previously observed range. The equivalent widths of H$\alpha$ in the two ALFOSC spectra are 40 and 70 $\AA$, similar to the full range of  45 to 96 $\AA$ in our spectra from 1996-1999.  

The most significant differences between spectra from the normal and faint states are found in the profiles of the resonance lines of \ion{Ca}{ii} and \ion{Na}{i} D and the forbidden [\ion{O}{i}] lines, as can be seen from the comparisons of selected lines in Figs. 2--5. Some of the details seen in the Sofin spectra are not resolved in the ALFOSC spectra. Nevertheless, one notices that the \ion{Na}{i} D lines, and especially the \ion{Ca}{ii} K line, had developed an enhanced wind feature in 2014 with P Cyg type profiles.

The [\ion{O}{i}] 6300 $\AA$ line is present in the high-resolution spectra obtained at CrAO and in the low resolution ALFOSC spectrum (Fig.~\ref{6300}). In our observations from 1996, and in earlier published observations (Hamann \cite{Hamann94}; Hartigan et al. \cite{Hartigan95}) the [\ion{O}{i}]  line has three emission components: a central one and two at radial velocities of about -140 and +140   km\,s$^{-1}$, which form in the two flows of a collimated wind (jet) at large distances from the star. In a series of spectra from 1999 presented in Alencar et al. (\cite{Alencar05}) the two jet components are at -170 and +120  km\,s$^{-1}$. In the 2014 spectra the two components appear at much lower velocities, namely about -90 and +90  km\,s$^{-1}$.

In addition, the equivalent widths of the forbidden lines of [\ion{O}{i}] and [\ion{S}{ii}] have changed drastically and are several times stronger as compared to the bright state. Considering that also the intensity of the stellar continuum is lower, we conclude that the {\it flux} in the lines did not change equally much. This implies that the occulting body covers the star and its immediate surrounding but does not extend further out to cover also the extended wind. The asymmetry in intensity of the blue- and red-shifted emission components,  seen in the CrAO spectrum of 2014, was present also in 1999 (Alencar et al.~({\cite{Alencar05}).

Finally, we did not find any trace of additional absorption components in the resonance \ion{Na}{i} and \ion{K}{i} lines, which would indicate a substantial increase of column density of cold gas toward the dimmed star. With the resolution of ALFOSC an absorption line with EW  $>$ 0.2 $\AA$ can be detected. Such components can hide in the complicated \ion{Na}{i} profiles, but the \ion{K}{i} line is not blended. However, in the ALFOSC spectrum the \ion{K}{i} absorption with EW = 0.4 $\AA$ is at -75 km\,s$^{-1}$ and must be related to the outflow. In the bright state the absorption of \ion{K}{i} is at stellar velocity, showing moderate night-to-night variability (Fig.~\ref{KI}).

\begin{figure}
\centerline{\resizebox{8.5cm}{!}{\includegraphics{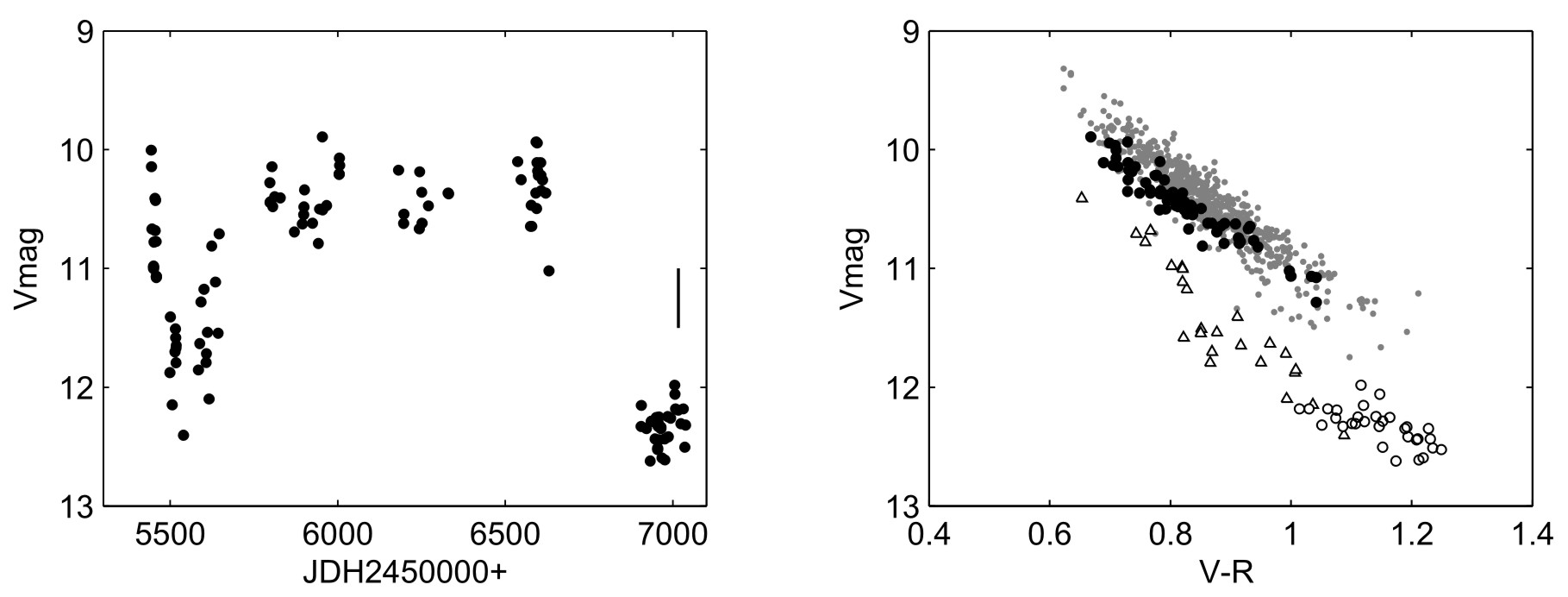}}}
\caption{ Left panel: light curve of RW Aur with the two minima during 2010 and 2014. The vertical bar indicates the time of our spectroscopic observations. Right panel: colour-magnitude diagram of RW Aur. Open circles -- minimum during 2014. Triangles -- minimum during 2010. Filled circles -- bright state between the two minima. Grey dots -- observations from 1986--2005. The colour index is in the Johnson system. }
\label{curve}
\end{figure} 

\begin{figure}
\centerline{\resizebox{8.5cm}{!}{\includegraphics{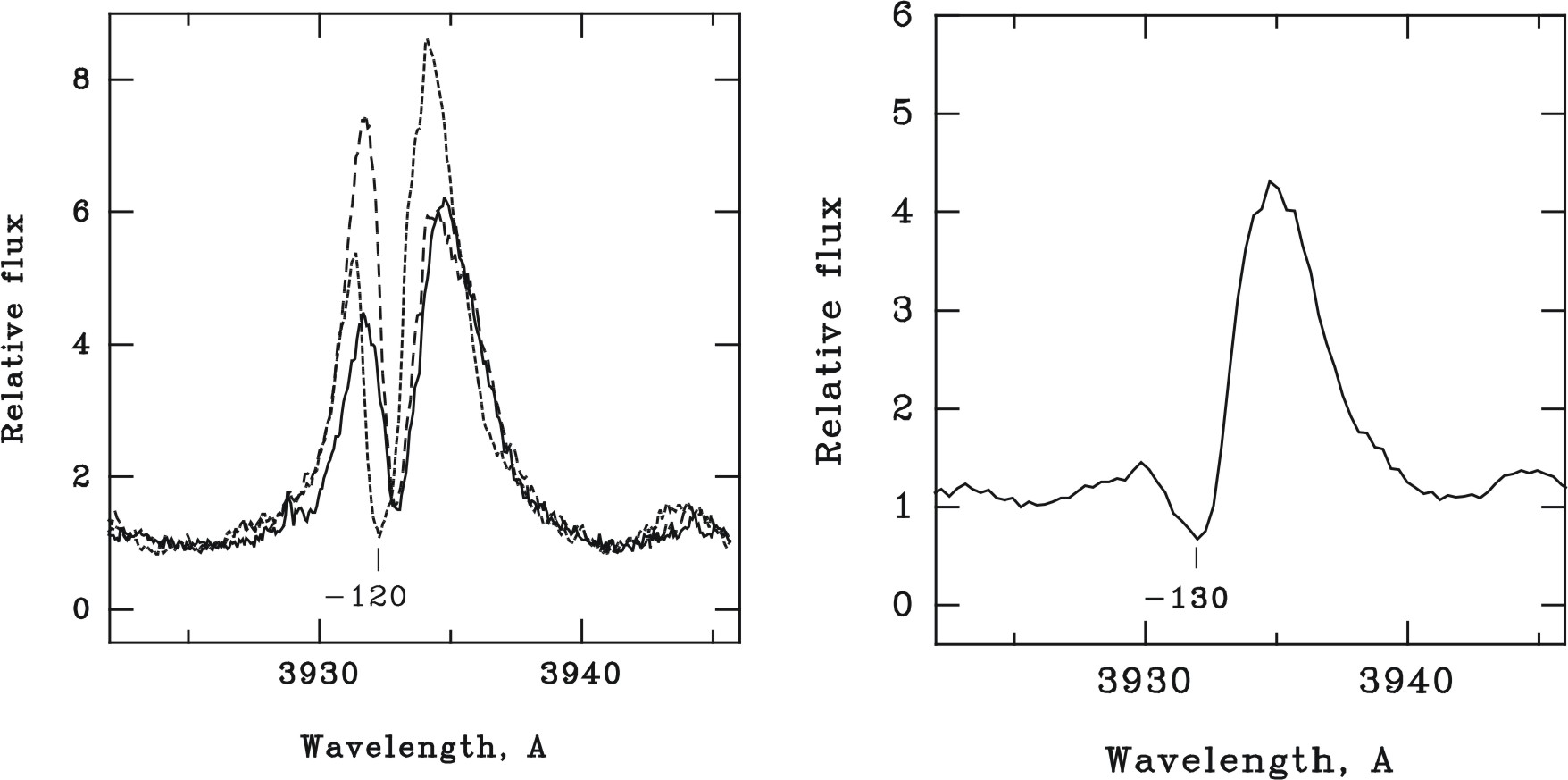}}}
\caption{Three spectra obtained during the bright state in 1998  covering the range of typical variations in the \ion{Ca}{ii} K line (left) and one corresponding spectrum in 2014 (right). Note that in 2014 the profile is of P Cyg type indicating an increased wind feature centred at -130 km\,s$^{-1}$. Here, and in the following figures the wavelength scale is heliocentric and velocities are in km\,s$^{-1}$ with respect to the star.}
\label{hal}
\end{figure} 

\begin{figure}
\centerline{\resizebox{8.5cm}{!}{\includegraphics{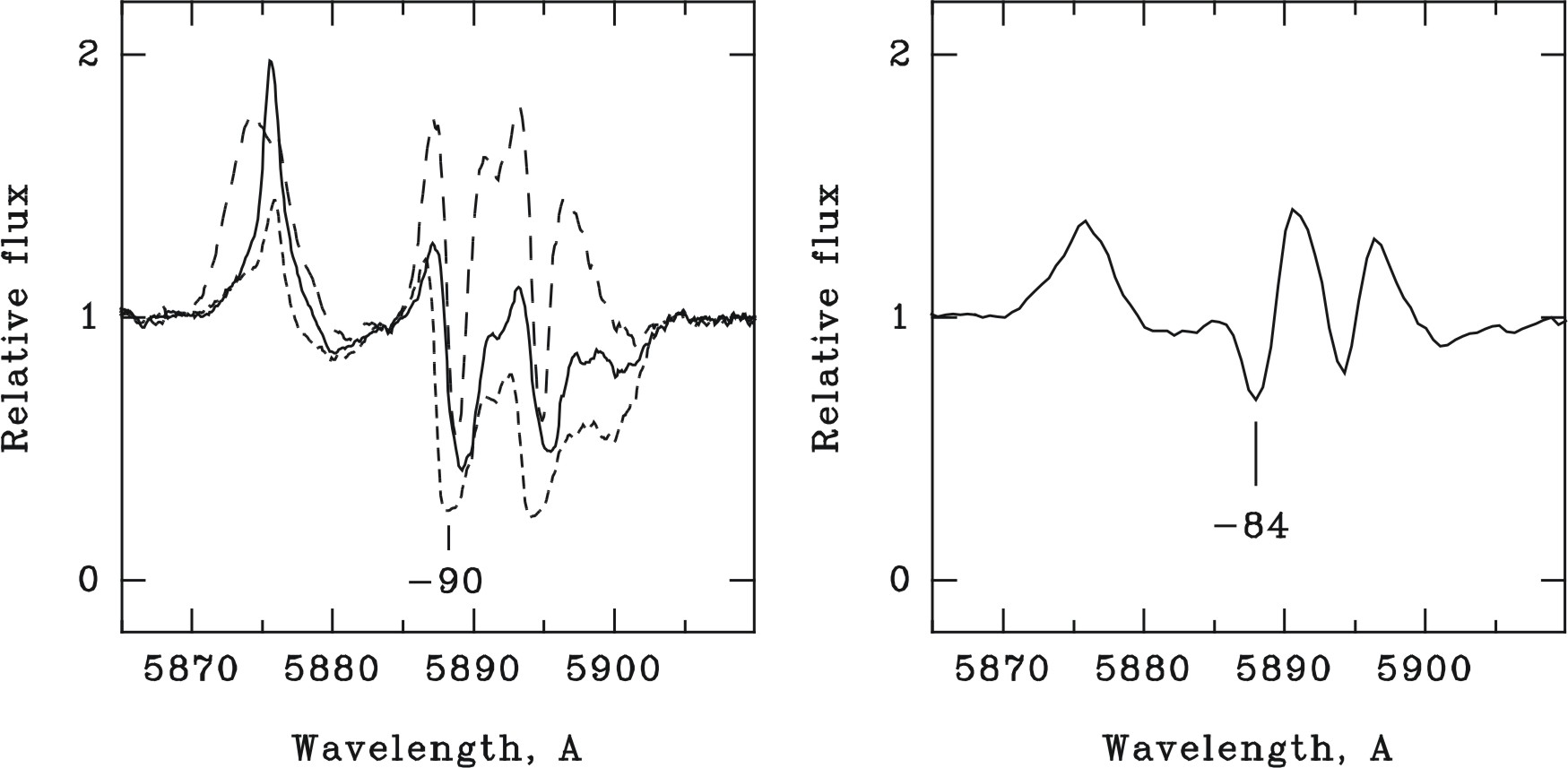}}}
\caption{Three spectra obtained during the bright state in 1996 showing typical variations in the \ion{He}{i} and \ion{Na}{i} D lines, and one corresponding spectrum in 2014 (right).} 
\label{dna}
\end{figure} 

\begin{figure}
\centerline{\resizebox{8.5cm}{!}{\includegraphics{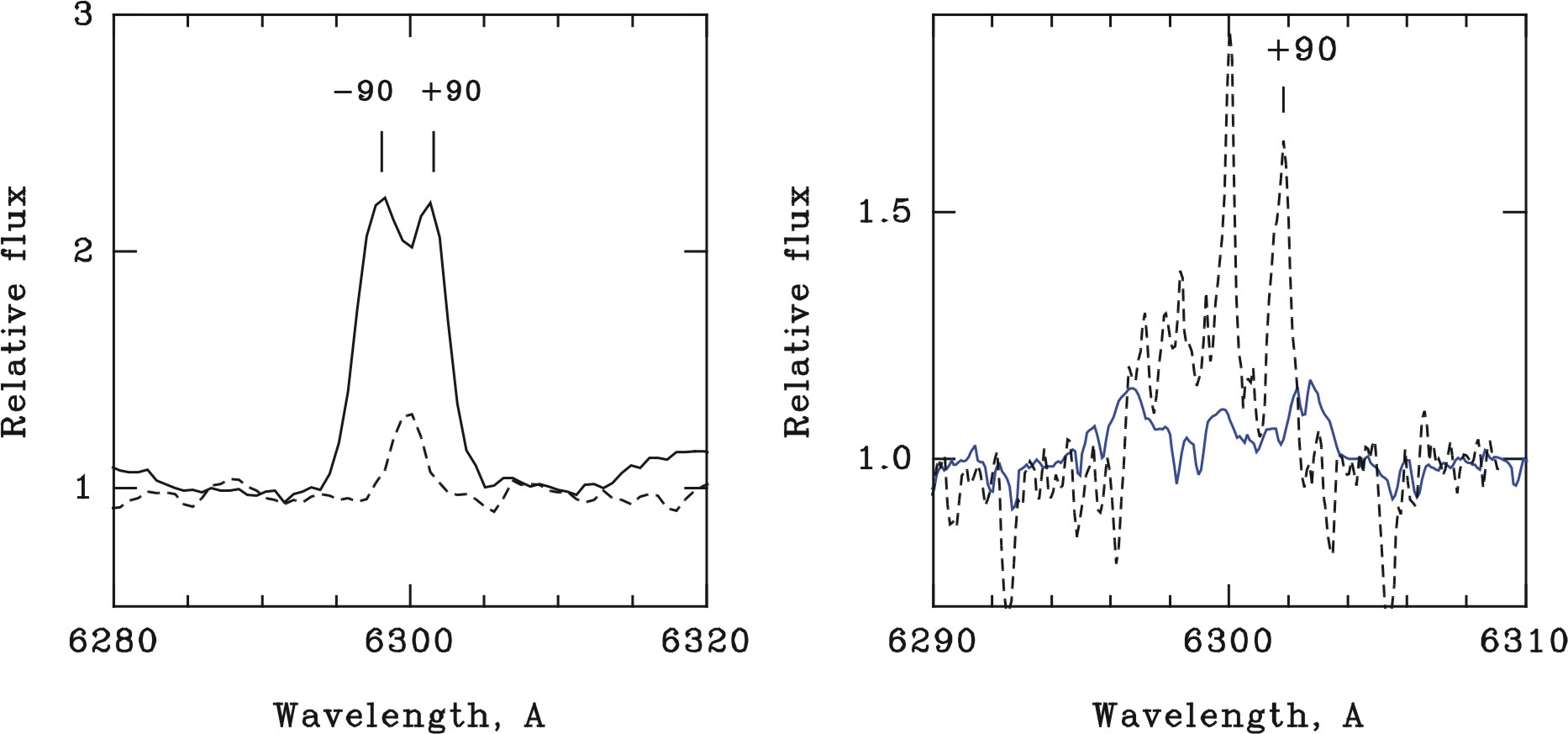}}}
\caption{Left panel -- ALFOSC spectra from 2014 covering the [\ion{O}{i}] 6300 line. The strong lines at -90 and +90 km\,s$^{-1}$ (solid) are from RW Aur A, while the single emission component at zero velocity (dashed) is from RW Aur B. Right panel --  high resolution spectra of the same region. Dashed curve shows the [\ion{O}{i}] profile from both RW Aur A and B in
2014, where the central peak belongs to RW Aur B, and the red-shifted sharp peak at +90 km\,s$^{-1}$ belongs to RW Aur A. The blue-shifted emission is shallow. Numerous narrow telluric absorption components and photospheric
lines from component B are present. The solid curve in the right panel shows the [\ion{O}{i}] profile in 1996, with the central peak at zero velocity and two peaks at -140 and + 140 km\,s$^{-1}$. The forbidden lines are much stronger in the faint than in the bright state.}
\label{6300}
\end{figure} 

\begin{figure}
\centerline{\resizebox{8.3cm}{!}{\includegraphics{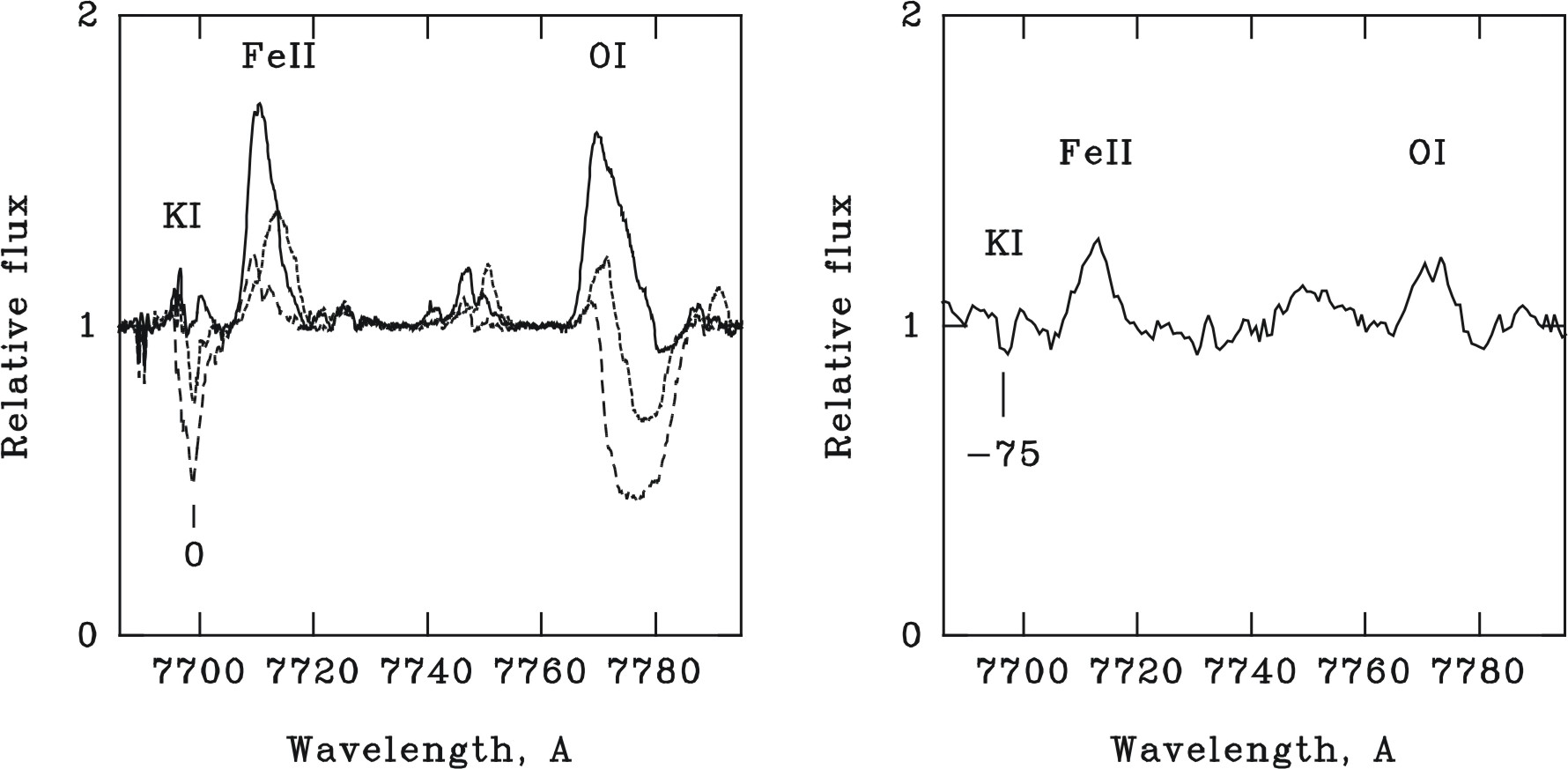}}}
\caption{The region covering the resonance \ion{K}{i} 7699 line and the \ion{O}{i} 7773 triplet. Left panel -- bright state in 1996, right panel -- faint state in 2014. Note the absence of any enhanced  \ion{K}{i} absorption components in the spectrum of the dimmed star; there is only a relatively weak absorption at -75  km\,s$^{-1}$.}
\label{KI}
\end{figure}

\section{Discussion}
\label{sec:discussion}

From our study of the spectral state of RW Aur A before and after the fading we can conclude that there are no changes in the characteristic emission line spectrum from He I, H and once and twice ionized metals. Therefore, there are no signs that the accretion flows close to the star has changed. On the other hand, we found that some drastic changes had occurred in the strength of forbidden lines of [\ion{O}{i}] and [\ion{S}{ii}], and in radial velocity of the two components of [\ion{O}{i}] 6300  formed in the bipolar jets of RW Aur A. In addition, an increase in the strength of wind signatures occurred in the resonance lines of \ion{Ca}{ii} and \ion{Na}{i}.

The increased intensity of the forbidden lines in RW Aur A, measured in units of the stellar continuum intensity, indicates that the observed flux in the lines remained about the same after the star dropped in brightness. This implies that the occulting body covered the star and its inner accretion flow, where the broad emission lines of metals form, and also the inner part of the stellar wind radiating in H$\alpha$. However, the more extended parts of the wind (the jets) were not occulted much. This sets certain restrictions to the size of the obscuring cloud.  

The photometry shows that the extinction becomes very grey when the star fades implying that large dust grains must be present in the obscuring body. The large extinction of Av $\sim$ 3$^m$ in combination with the absence of additional absorption lines in resonance lines of \ion{Na}{i} and \ion{K}{i} indicate that the dust-to-gas ratio in the obscuring cloud could be very different from normal conditions in the interstellar medium.

The dimmings of RW Aur A are of a different  nature than those observed in  AA Tau and KH 15D (see Sect.~\ref{sec:intro}). In the hypothesis proposed in R2013 the occultation of RW Aur A in 2010 was caused by a dusty structure connected to a tidally disrupted disk. Since the phenomenon has returned a second time one must assume that in this case the foreground structure contains more than one condensation passing in front of the star. 

An alternative explanation to explore is whether the dimming is caused by some kind of interaction between a changing outflow and the disk leading to a transport of larger dust grains from the diskplane into the line-of-sight far above the inclined disk. The problem of dust survival in disk winds of CTTS was investigated by Tambovtseva \& Grinin (\cite{Tambovtseva08}). The emergence  of dust into the wind can be caused either by disk erosion (e.g. Schnepf et al.\cite{Schnepf14}), or by the MHD processes at the disk-magnetosphere boundary (e.g., review by Romanova et al. \cite{Romanova13}). In this case the star and the H$\alpha$-emitting region are obscured by a screen related to a dusty wind containing large dust grains.

The remarkable phenomenon of the long-lasting deep minima in RW Aur A rises a number of questions. How recurrent are these occultations? Are there any similarities to UX Ori stars? Are the dimmings in fact related to a changing stellar wind? It is premature to draw any deeper conclusions about the cause of the phenomenon. Continued spectroscopic, {photometric and polarimetric} observations of the star will provide more insight, and it will be interesting to see whether the normal  wind features are restored when RW Aur A returns to its normal state. 

\section{Conclusions}
\label{sec:conclusion}

We have observed the binary T Tauri star RW Aur A and B spectroscopically and photometricly during the deep, long-lasting minimum that occurred on the primary in 2014. This is the second time over at least 50 years that RW Aur A suffers from such a dimming. Photometric data indicate that the star has been occulted by a foreground cloud containing large dust grains. Our primary concern has been to search for spectral changes that could be related to the phenomenon. 

We conclude that the unusually rich emission line spectrum of RW Aur A on the whole remained the same as at normal brightness levels. For instance, spectral signatures of accretion flows in the immediate surrounding of the star have not changed, and components in emission line profiles of H$\alpha$ that trace the inner part of a stellar wind/jet remained the same. 

On the other hand, significant changes occurred in line components related to more extended parts of the stellar wind, such as enhanced blue-shifted absorption in resonance lines from e.g. \ion{Ca}{ii}. Moreover, forbidden lines of e.g. [\ion{O}{i}] and [\ion{S}{ii}] became considerably stronger than in the bright state, and the radial velocities of the two components tracing more extended emission from the oppositely directed flows in the wind/jet were different. The change in equivalent width of these lines can be explained as caused by an occulting body that covers only the inner part of the stellar wind, and not the remote parts. 

The picture is not clear regarding the origin of the obscuring cloud containing large dust grains,
but one would expect that these grains once resided in the disk plane. Since the disk of RW Aur A is inclined by more than 30 degrees one must find ways to explain how this material was transported far above the disk, into the line-of-sight. One explanation has been that the occulting body is part of a tidally disrupted disk. Our results do not rule out this hypothesis, but since we have found evidence that the stellar wind has changed after the fading we speculate that an interaction between this enhanced wind and the dusty disk caused an ejection of large dust grains into the wind. We did not find any traces of absorption by cool gas related to the obscuring cloud.

\begin{acknowledgements}
Based on observations made with the Nordic Optical Telescope, operated by the Nordic Optical Telescope Scientific Association (NOTSA) at the Observatorio del Roque de los Muchachos, La Palma, Spain, of the Instituto de Astrofisica de Canarias. The data presented here were obtained with ALFOSC, which is provided by the Instituto de Astrofisica de Andalucia (IAA) under a joint agreement with the University of Copenhagen and NOTSA. KG acknowledges funding from the LabEx OSUG@2020 that allowed purchasing the imaging system installed at the 1.25-m telescope at CrAO. PP thanks Marina Romanova for usefull discussions.
  
\end{acknowledgements}

\end{document}